\documentclass[twocolumn, superscriptaddress,
 amsmath,amssymb,
 aps,prb,
 longbibliography,
]{revtex4-2}

\usepackage{graphicx}
\usepackage{dcolumn}
\usepackage{bm}
\usepackage{physics}
\usepackage{here}
\usepackage{xcolor}
\usepackage{braket}
\usepackage[
 colorlinks=true,
 allcolors=blue,
]{hyperref}

\usepackage{comment}
\usepackage{color}

\usepackage{mathdots}

\begin{document}
\title{Floquet-Weyl states at one-photon resonances in three-dimensional topological insulators}

\author{Keiya Uehara}
\email{keiya-uehara@g.ecc.u-tokyo.ac.jp }
\affiliation{Department of Applied Physics, The University of Tokyo, Tokyo 113-8656, Japan}
\affiliation{Department of Applied Physics, Tokyo University of Science, Tokyo 125-8585, Japan}

\author{Ryo Okugawa}
\affiliation{Department of Applied Physics, Tokyo University of Science, Tokyo 125-8585, Japan}

\author{Takami Tohyama}
\affiliation{Department of Applied Physics, Tokyo University of Science, Tokyo 125-8585, Japan}

\author{Shun Okumura}
\email{okumura@ap.t.u-tokyo.ac.jp}
\affiliation{Quantum-Phase Electronics Center (QPEC), The University of Tokyo, Bunkyo, Tokyo 113-8656, Japan}
\affiliation{RIKEN Center for Emergent Matter Science (CEMS), Wako, Saitama 351-0198, Japan}
\affiliation{WPI-SKCM$^2$, Hiroshima University, Higashi-Hiroshima, Hiroshima 739-8531, Japan}
 
%\date{\today}

\begin{abstract}
Quantum materials exhibit exotic phases and electronic responses under irradiation by circularly polarized light, which breaks time-reversal symmetry and generates Floquet replica bands. 
Recently, Floquet topological states arising from direct resonances have attracted much attention, e.g., the emergence of Floquet-Weyl points at a one-photon resonance, rather than topological features within the modulated original bands via high-frequency expansion.  
In this study, we investigate the effects of a one-photon resonance in a representative three-dimensional topological insulator, $\text{Bi}_2\text{Se}_3$, applying Floquet theory under circularly polarized light. 
We find that four pairs of Floquet-Weyl points emerge in the intermediate-frequency regime, mediated by hybridization between the original and one-photon-resonant Floquet bands, preserving the threefold rotational symmetry of the crystalline structure. 
Our numerical calculations demonstrate that tuning the chemical potential via hole doping yields a large anomalous Hall conductivity, directly associated with these Floquet-Weyl points. 
This work provides a highly accessible route toward the experimental realization of one-photon-resonant Floquet-Weyl semimetals.
\end{abstract}

\maketitle

\section{INTRODUCTION}
Floquet engineering provides a powerful framework for controlling nonequilibrium states of quantum materials by time-periodic external fields~\cite{BukovAP2015,basovNM2017,OkaARCMP2019,rudnerNRP2020,TorreRMP2021}.
A central idea is that coherent optical driving can dress electronic states and modify their band topology on ultrafast time scales.
Much of the early theoretical progress was based on the high-frequency, or off-resonant, regime, where the photon energy is larger than the relevant electronic energy scales.
In this limit, the time-periodic problem can be assigned to a time-independent effective Hamiltonian through the Floquet--Magnus expansion or related high-frequency expansions (HFE)~\cite{magnusCPA1954,BlanesPR2009,GoldmanPRX2015,EckardtIOP2015,MikamiPRB2016}.
This approach has provided a broad design principle for photo-induced effective gauge fields, Berry curvature, and topological phases. 
In two-dimensional (2D) systems, for a typical example, circularly polarized light (CPL) induces a Hall response in graphene~\cite{OkaPRB2009}, and leads to photo-induced quantum Hall insulators without Landau levels~\cite{KitagawaPRB2011}. 
The related Floquet topological insulator (TI) with helical edge states was proposed in semiconductor quantum wells~\cite{lindnerNP2011}. 
These theoretical developments, together with synthetic realizations in cold atoms and photonic systems~\cite{RechtsmanNat2013,jotzuNat2014}, established high-frequency Floquet engineering as a major route to nonequilibrium topological states.

Beyond HFE in the off-resonant regime, photon-resonance effects have recently attracted increasing attention. 
In realistic laser experiments, the photon energy is often comparable to interband transition energies, where hybridization between the original and Floquet-replica bands becomes essential. 
On the theoretical side, pump-probe spectra of graphene have been analyzed in terms of Floquet-band formation and local pseudospin textures~\cite{SentefNC2015}, while resonant gap opening and nonequilibrium topological transitions have been discussed in driven 2D systems~\cite{IadecolaPRB2014,kunduPRL2014}. 
The corresponding boundary physics has also been explored through Floquet chiral edge states in graphene~\cite{perezPRB2014}. 
In addition, nonequilibrium occupations, dissipation, and reservoir effects have been shown to play crucial roles in determining the observable Hall response of driven systems~\cite{seetharamPRX2015,dehghaniPRB2015,satoPRB2019}. 
In graphene, experimentally, the light-induced anomalous Hall effect (AHE) due to the Floquet topological band has been reported under the CPL~\cite{mciverNP2020}, and recent time-resolved spectroscopies have revealed direct evidence of Floquet states and Floquet-induced hybridization gaps~\cite{merboldtNP2025,wangNM2026}. 
These developments show that resonant Floquet physics is not merely a correction to the off-resonant HFE limit, but provides a distinct mechanism for controlling nonequilibrium band topology.

Floquet engineering has also been extended to a variety of three-dimensional (3D) electronic structures, such as 3D Dirac semimetals~\cite{narayanPRB2015,ebiharaPRB2016,hubenerNat2017}, Weyl semimetals~\cite{chanPRL2016,taguchiPRB2016}, nodal-line semimetals~\cite{yanPRL2016,chanPRB2016,chenPRB2018}, and semiconductors~\cite{zhangPRB2016}, where the optical control of Weyl points via HFE provides a clear route to nonequilibrium topological transport. 
Nevertheless, many of these approaches start from semimetallic band structures with Dirac, Weyl, or nodal-line crossings and require light-induced band splitting large enough to compete with an intrinsic gap. 
Recent studies of light-induced Hall responses in massive 3D Dirac semimetals have further highlighted the need to distinguish intrinsic Floquet-band contributions from other nonequilibrium mechanisms~\cite{MurotaniPRL2023,YoshikawaPRB2025}. 
On the other hand, in the lower-frequency regime, resonant hybridization between Floquet bands can generate Weyl points accompanied by Fermi-arc surface states~\cite{BucciantiniPRB2017}, and optically resonant interband transitions in 3D semiconductors have also been proposed as a mechanism for Floquet-Weyl semimetals~\cite{zhangPRB2022}. 
In particular, a recent proposal by Hirai \textit{et al.}~\cite{HiraiPRR2024} showed that Weyl points can emerge at one-photon resonance in 3D Dirac electrons, providing a generic way for the realization of Floquet-Weyl states without relying on HFE. 

In this paper, motivated by this one-photon resonance mechanism, we investigate Floquet-Weyl generation driven by a CPL in representative 3DTIs, Bi$_2$Se$_3$-type compounds~\cite{XiaNatPhys2009} [Fig.~\ref{fig:1}(a)]. 
Experimentally, Floquet--Bloch bands have been observed on TI surfaces~\cite{WangSci2013}, and their ultrafast build-up and dephasing have been resolved on subcycle time scales~\cite{ItoNat2023}. 
On the theoretical side, while Bi$_2$Se$_3$-type 3DTIs have been proposed as platforms for off-resonant Floquet-Weyl phases~\cite{wangIOP2014}, the resonance effect on Floquet-Weyl generation remains unexplored.
We show that a CPL generates Floquet-Weyl points through the resonant hybridization between the original valence and photon-shifted conduction bands, as shown in Fig.~\ref{fig:1}(b). 
The resulting configuration of the Weyl points is strongly constrained by the crystalline symmetry of Bi$_2$Se$_3$, leading to a characteristic threefold splitting pattern beyond the description of the isotropic continuum limit. 
By tracking the creation, motion, and annihilation of these Weyl points, we construct a topological phase diagram containing gapped, Floquet-Weyl, and Floquet-Chern states. 
We further calculate the photo-induced anomalous Hall conductivity (AHC) and show that hole doping can shift the chemical potential toward the resonant energy level, thereby revealing a sharp peak in AHC originating from the Weyl points. 
Our results establish one-photon resonance as a versatile way to design Floquet-Weyl states in spin--orbit-coupled quantum materials beyond conventional 3D Dirac semimetals.

The organization of this paper is as follows. 
In Sec.~\ref{sec:model}, we introduce the model Hamiltonian, starting with the tight-binding model in equilibrium, followed by the formulation of the one-photon resonant Floquet Hamiltonian.
In Sec.~\ref{sec:band}, we discuss the Floquet band structures, elucidate the formation of Weyl points at the one-photon resonance, and present the calculated topological phase diagram and the trajectories of Weyl points. 
In Sec.~\ref{sec:transport}, we investigate the macroscopic electronic transport properties, specifically focusing on the photo-induced AHC, and demonstrate that hole doping can effectively extract the topological responses originating from the emergent Weyl points. 
Finally, Section~\ref{sec:conclusion} is devoted to a summary of our findings and perspectives.

\section{MODEL}
\label{sec:model}

In this section, we construct a theoretical framework to investigate the topological and transport properties of a representative 3DTI Bi$_2$Se$_3$ under irradiation of a CPL. 
We start from a tight-binding model that mimics the equilibrium electronic structure of Bi$_2$Se$_3$ in Sec.~\ref{sec:model_tb}, and we employ the Floquet formalism with a one-photon approximation in Sec.~\ref{sec:model_opr}.

\subsection{Tight-binding model in equilibrium}
\label{sec:model_tb}

\begin{figure}[t]
    \centering
    \includegraphics[width=1.0\linewidth]{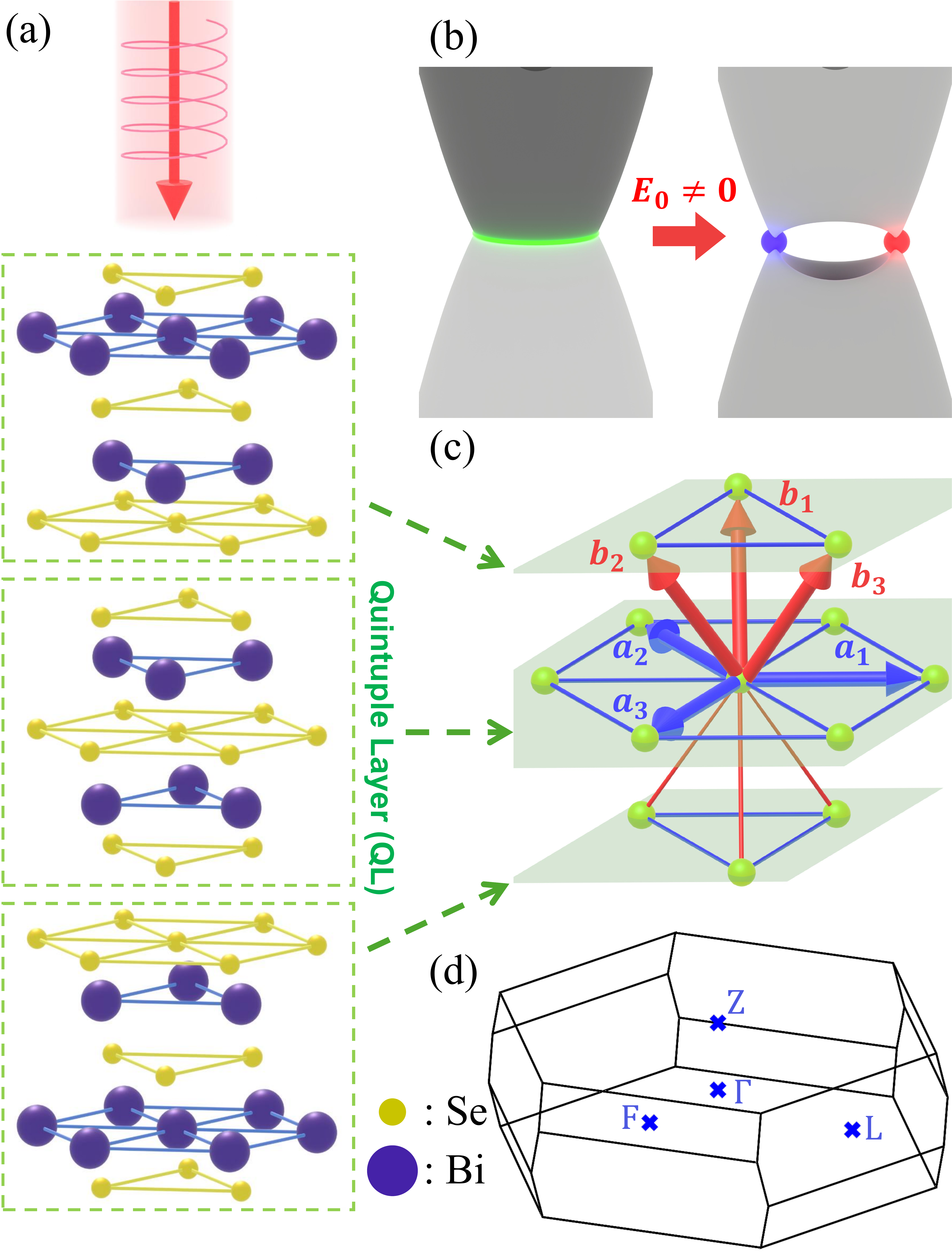}
    \caption{Schematics of (a) the crystal structure of $\rm{Bi_2Se_3}$ irradiated by the circularly polarized light, (b) a nodal ring between the original and Floquet-replica bands, turning to the Weyl points with a finite electric field $E_0$, and (c) the simplified triangular lattice model.
    The dashed boxes in (a) represent the units of quintuple layers, which are reduced to the green sheets in (c).
    In (c), the green spheres denote each unit cell, and the blue (red) arrows are the intra-(inter-)layer translational vectors $\mathbf{a}_l$ ($\mathbf{b}_l$).
    (d) The first Brillouin zone of the rhombohedral lattice. 
    The four inequivalent time-reversal-invariant momenta (TRIM) are marked by blue crosses: $\Gamma, \text{Z}, \text{F}$, and $\text{L}$.
    }
    \label{fig:1}
\end{figure}

To describe the low-energy electronic structure of Bi$_2$Se$_3$, we employ the effective tight-binding model~\cite{zhangNat2009,LiuPRB2010,MaoPRB2011}.
The model is based on the rhombohedral crystal structure with space group $R\bar{3}m$, in which the basic structural unit is a quintuple layer (QL) consisting of atomic layers of Se-Bi-Se-Bi-Se, as shown in Fig.~\ref{fig:1}(a).
Since the intra-QL bonding is much stronger than the inter-QL van der Waals coupling, each QL can be regarded as an effective triangular-lattice layer stacked along the $z$ direction, as shown in Fig.~\ref{fig:1}(c).
Starting from the Bi and Se $p$ orbitals near the $\Gamma$ point, the strong chemical bonding within a QL first hybridizes these orbitals into parity eigenstates. 
Subsequently, the crystal-field splitting separates the $p_z$-derived states from the $p_x$ and $p_y$ states, leaving two opposite-parity states near the Fermi level.
Finally, strong spin--orbit coupling (SOC) induces a level repulsion between them, and realizes the inverted band structure.
Following the notation in Ref.~\cite{MaoPRB2011}, the relevant low-energy degrees of freedom are reduced to four spinful states, $\ket{P_z^\pm,\sigma}$, where $\pm$ and $\sigma = \uparrow$ or $\downarrow$ denote the parity and spin of the state, respectively.
These four states provide a minimal set of bases for the conduction ($\ket{P_z^-,\sigma}$) and valence ($\ket{P_z^+,\sigma}$) bands, which capture the essential topological properties and low-energy electronic structure of Bi$_2$Se$_3$-type 3DTIs around the $\Gamma$ point.

Consequently, we can construct an effective tight-binding model on the stacked triangular lattices with the intra- and inter-layer hopping terms, whose translational vectors are denoted by $\mathbf{a}_l$ and $\mathbf{b}_l$ in Fig.~\ref{fig:1}(c), respectively.
The corresponding first Brillouin zone (BZ) is plotted in Fig.~\ref{fig:1}(d) with the high-symmetry points. 
This model possesses twofold(threefold) rotational symmetry on the $x$-($z$-) axis, $C_{2x}$($C_{3z}$), in addition to the spatial inversion symmetry $\mathcal{P}$ and the time reversal symmetry $\mathcal{T}$~\cite{MaoPRB2011}. 
In momentum space, the effective Bloch Hamiltonian is given by
\begin{align}
    \mathcal{H}(\mathbf{k}) &= h_0(\mathbf{k}) \mathbb{I}_4 + \sum_{i=1}^5 h_i(\mathbf{k}) \Gamma_i,
\label{eq:H_tb}
\end{align}
where $\mathbf{k}$ represents the 3D vector of the wavenumber, $\mathbb{I}_n$ is the $n\times n$ identity matrix, and $\Gamma_i$ matrices are the direct products of the Pauli matrices $\sigma_i$ and $\tau_i$ for the spin and parity degrees of freedom, respectively; $\Gamma_i = \sigma_i\otimes\tau_1$ for $i=1,2,3$, $\Gamma_4 = \mathbb{I}_2\otimes\tau_2$, and $\Gamma_5 = \mathbb{I}_2\otimes\tau_3$.

The coefficients $h_i(\mathbf{k})$ in Eq.~\eqref{eq:H_tb} include the real parameters, $A_0 (B_0)$ and $A_{nm}(B_{nm})$, for intra-(inter-)layer hopping, where the suffixes $0$ and $nm$ mean the scalar and $(n,m)$ elements of the $4\times4$ matrices, respectively.
In the following calculations, we use a set of these model parameters and the Fermi energy $E_\mathrm{F}$ numerically optimized by fitting the band structure to the {\it ab-initio} data for Bi$_2$Se$_3$ obtained from Wannier Tools~\cite{WannierTools2018}.
The optimization is performed to achieve the best fit near the $\Gamma$ point while keeping the intra-layer hopping larger than the inter-layer hopping, namely, $A_0 > B_0$ and $A_{nm} > B_{nm}$, corresponding to the parameter regime of the strong TI.
The detailed expression of the translational vectors $\mathbf{a}_l$ and $\mathbf{b_l}$, the coefficients $h_i(\mathbf{k})$, and model parameters $A_0 (B_0)$ and $A_{nm}(B_{nm})$ are provided in Appendix~\ref{ap:model}.

\subsection{One-photon-resonant Floquet Hamiltonian}
\label{sec:model_opr}

To address the nonequilibrium steady state of the 3DTI, we adopt the Floquet theory in the system periodically driven by the CPL. 
We incorporate the effect of the CPL through the Peierls substitution, $\mathbf{k}\rightarrow \mathbf{k} + \mathbf{A}(t)$, where the right-handed CPL along the $-z$ direction is introduced by the vector potential $\mathbf{A}(t)=(\frac{E_0}{\Omega}\cos{\Omega t}, \frac{E_0}{\Omega}\sin{\Omega t}, 0)$; $E_0$ and $\Omega(>0)$ are the amplitude of the electric field and the photon energy, respectively.
Here we take the elementary charge $e=1$ and the reduced Planck constant $\hbar = 1$.
Performing Fourier transformations in the time-periodic Hamiltonian $H(t) = H(t+T)$ with the periodicity $T = \frac{2\pi}{\Omega}$, we obtain the full Floquet Hamiltonian $H_\mathrm{F}$ in the matrix form, written as 
\begin{equation}
      \label{eq:Floquet_matrix}
      H_\mathrm{F} = 
\begin{pmatrix}
  \ddots  & \vdots & \vdots & \vdots & \iddots \\
  \dots & H_0 + \Omega & H_1 & H_2  & \dots \\
  \dots & H_{-1} & H_0 & H_1  & \dots \\
  \dots & H_{-2} & H_{-1} & H_0 - \Omega  & \dots \\
  \iddots & \vdots & \vdots & \vdots & \ddots
\end{pmatrix},
\end{equation}
where $H_m = \frac{1}{T}\int^T_0dtH(t)e^{-im\Omega t}$.
The resulting full Floquet Hamiltonian contains infinitely many photon sectors and therefore has to be truncated in practical calculations.

To capture the band hybridization responsible for the generation of Weyl points, we focus on the minimal subspace relevant to the one-photon-resonant state~\cite{HiraiPRR2024}.
The relevant one-photon resonance is dominated by hybridization between the $0$-photon and the $-1$-photon sectors, which are originally associated with the valence and conduction bands in equilibrium, respectively.
The effective Floquet Hamiltonian for the one-photon-resonant state $\mathcal{H}_\mathrm{eff}$ is given by the $8\times8$ matrix, 
\begin{equation}
    \mathcal{H}_{\mathrm{eff}} = 
    \begin{pmatrix}
        H_0 & H_{+1}\\
        H_{-1} & H_0 - \Omega
    \end{pmatrix},
    \label{eq:H_opr}
\end{equation}
where $H_0$ describes the time-averaged Hamiltonian, whereas $H_{\pm1}$ describes the one-photon absorption and emission processes induced by the CPL, satisfying $H_{+1} = H_{-1}^\dagger$. 
The detailed expressions of the Floquet Hamiltonian will be provided in Appendix~\ref{ap:model}.

\section{Electronic band topology}
\label{sec:band}

\subsection{Band structures}
\label{sec:band_}

\begin{figure}[b]
    \centering
    \includegraphics[width=1.0\linewidth]{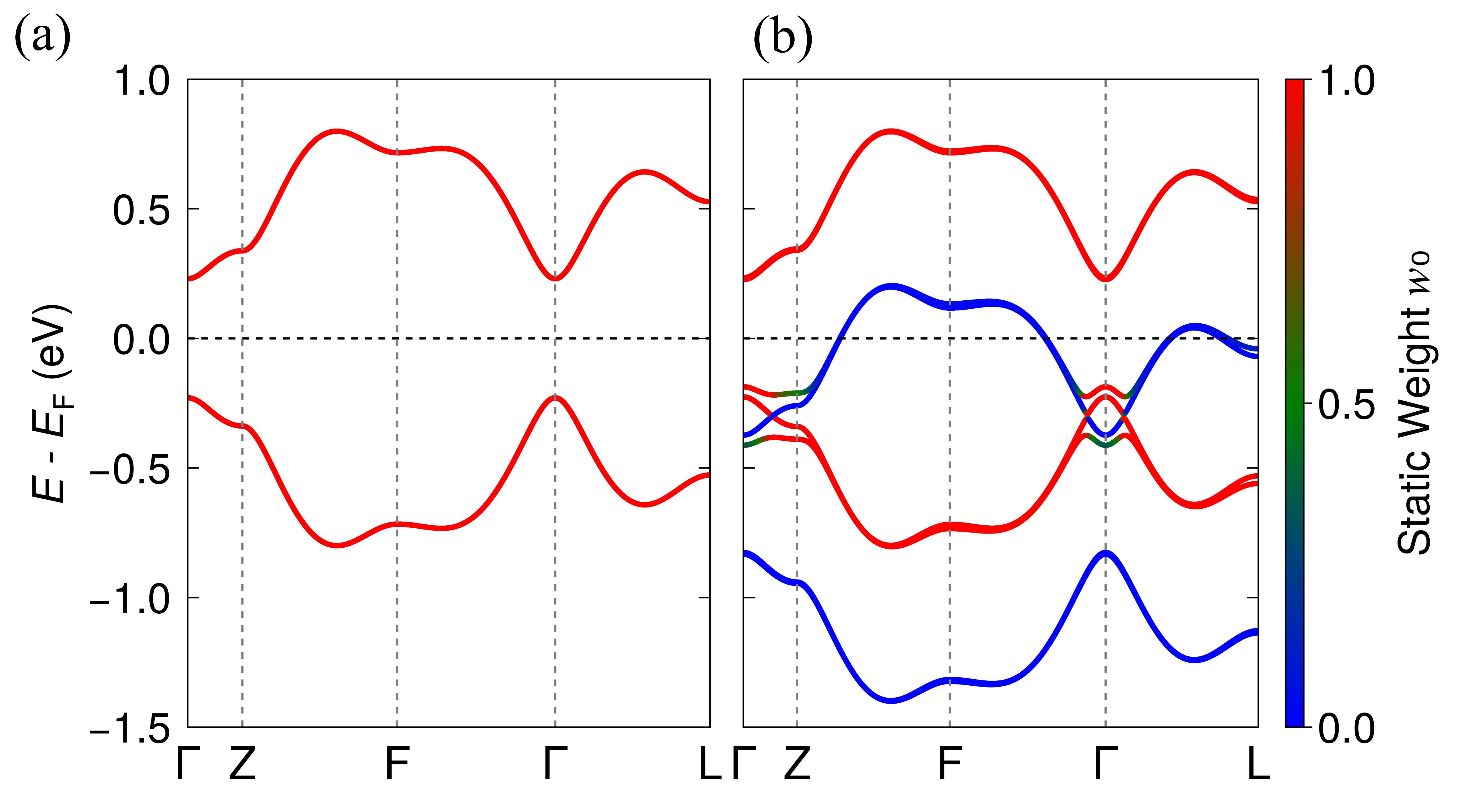}
    \caption{Band structures of Bi$_2$Se$_3$ calculated from (a) the tight-binding model in equilibrium and (b) the one-photon-resonant model driven by the circularly polarized light with photon energy $\Omega = 0.6$~eV and an electric field amplitude $E_0 = 2.4$~MV/cm. 
    The color scale encodes the static weight $w_0$, in which $w_0 = 1$ and $0$ correspond to the $0$- and $-1$-photon states, respectively.}
    \label{fig:2}
\end{figure}

We first examine the equilibrium band structure of the four-band tight-binding model in Eq.~\eqref{eq:H_tb} using exact diagonalization.
Figure~\ref{fig:2}(a) shows the energy dispersion from the Fermi level, $E - E_\mathrm{F}$, along the high-symmetry lines of the BZ in Fig.~\ref{fig:1}(d).
At the $\Gamma$ point, a direct gap of $\sim 0.46$~eV opens between the spin-degenerate conduction and valence bands due to the strong SOC in the presence of time-reversal and inversion symmetries, as discussed in Sec.~\ref{sec:model_tb}.
The fitted band structure is nearly particle-hole symmetric around $E_F$, since the particle-hole-asymmetric terms proportional to $A_0$ and $B_0$ are negligible to the other hopping parameters.

Turning on the CPL, we next plot the Floquet band structure in the one-photon-resonant model in Eq.~\eqref{eq:H_opr}.
In the absence of the electric-field amplitude $E_0$, the Floquet spectrum consists of replicas of the equilibrium bands shifted by the photon energy $\Omega$; therefore, the $-1$-photon sector appears as a copy of the static bands shifted downward by $\Omega$.
Once $E_0$ becomes finite, the $0$- and $-1$-photon sectors hybridize through the light-induced coupling $H_{\pm1}$.
Figure~\ref{fig:2}(b) exhibits the eight Floquet bands for $\Omega = 0.6$~eV and $E_0 = 2.4$~MV/cm.
The $n$th Floquet band is colored by the static weight $w_0 = \bra{n}\frac{1}{2}(\eta_z+\mathbb{I}_{2\times2})\otimes\mathbb{I}_{4\times 4}\ket{n}$, where $\ket{n}$ is the $n$th eigenstate and $\eta_z$ represents the Pauli matrix for the photon-number degree of freedom.
The band modulation is most pronounced near the $\Gamma$ and $\mathrm{Z}$ points, where the valence and conduction bands strongly hybridize at the one-photon resonance in $E - E_\mathrm{F} \sim -\frac{\Omega}{2}$. 
Despite this resonant hybridization, the band crossings appear in the quasienergy spectrum, which are responsible for the emergence of Floquet-Weyl points as introduced in the following subsection. 
By contrast, the overall band deformation, in particular prominent around the $\mathrm{L}$ and $\mathrm{F}$ points, is mainly governed by off-resonant virtual photon processes and can be understood within HFE, as discussed in Appendix~\ref{ap:HFE}.

\subsection{Floquet-Weyl points at one-photon resonance}

\begin{figure}[t]
    \centering
    \includegraphics[width=1.0\linewidth]{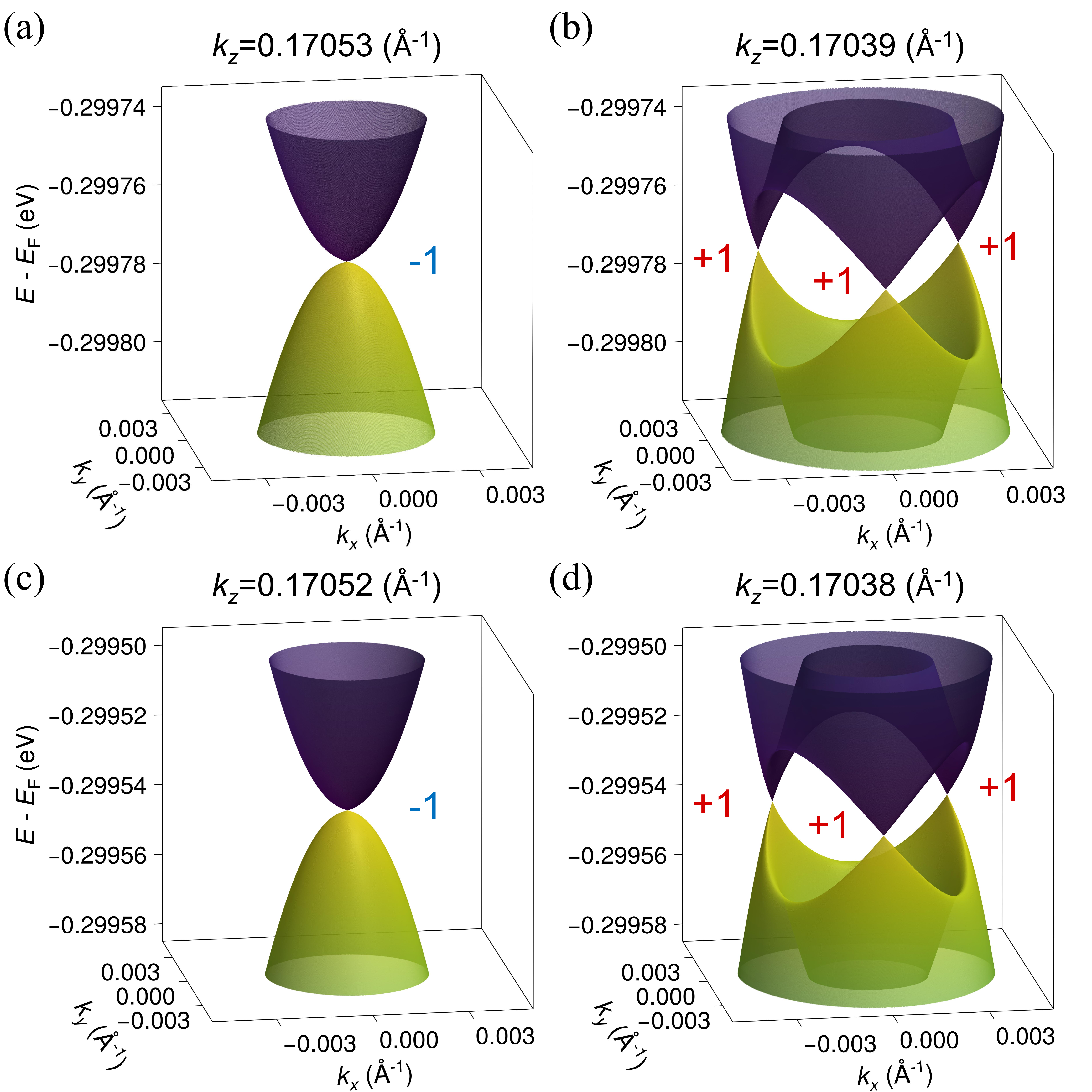}
    \caption{Energy dispersion obtained from (a)(b) the one-photon-resonant model in Eq.~\eqref{eq:H_opr}, corresponding to the fourth and fifth bands in Fig.~\ref{fig:2}(b), and (c)(d) the effective two-band model in Eq.~\eqref{eq:2by2H} for slightly different $k_z (> 0)$ around the Weyl points.
    The Weyl point on the $\Gamma$-$\mathrm{Z}$ line in (a) and (c) has the negative topological charge $-1$, while the three Weyl points around the $\Gamma$ point in (b) and (d) have the positive charge $+1$.}
    \label{fig:3}
\end{figure}

In order to clarify the topological natures of the one-photon-resonant states, we focus on the band crossing between the fourth and fifth Floquet bands in Fig.~\ref{fig:2}(b).
Along the $\Gamma$-$\mathrm{Z}$ line, we numerically find a gapless node with the Weyl charge $-1$ at $k_z \sim 0.17053$ as shown in Fig.~\ref{fig:3}~(a).
The topological charge is evaluated by using the discretized Berry flux method~\cite{FHS_method, HirayamaJPSJ2018}.
In practice, we construct a small cubic closed surface enclosing the gapless nodal point and integrate the Berry flux over this manifold to obtain the quantized topological charge.
Furthermore, as shown in Fig.~\ref{fig:3}~(b), we also find three other Weyl points with each Weyl charge $+1$ on the $k_z \sim 0.17039$ plane, on the $\mathrm{F}$-$\mathrm{\Gamma}$-$\mathrm{L}$ line in Fig.~\ref{fig:2}(b).
In total, eight Weyl points are created in pairs preserving symmetries of the crystalline structure, i.e., the spatial inversion $\mathcal{P}$ and $C_{3z}$ symmetries. 
In this case, Weyl points related by spatial inversion have opposite topological charges~\cite{OkugawaPRB2014, armitageRMP2018}, while those related by $C_{3z}$ symmetry carry the same topological charge~\cite{MurakamiSA2017}.

Here, we explore the origin of the emergent Weyl points at the one-photon resonance. 
Figs.~\ref{fig:3}(c) and \ref{fig:3}(d) show plots of the effective two-band Hamiltonian obtained by projecting the $8\times8$ Floquet Hamiltonian onto degenerate resonant states that form the Weyl points. 
This effective two-band Hamiltonian $\mathcal{H}_{2\times2}$ is expressed as 
\begin{align}
    &\mathcal{H}_{2\times2} = \left( h_0(\bm{q})-\frac{\Omega}{2} \right) \mathbb{I}_2 \notag\\
    & +  
    \begin{pmatrix}
        -c_{1z}cq_z-c'_{2\perp}a^2q_+q_- && a^2A\left( \lambda_{1\tau}q_+ + a\lambda_{2\tau}q_-^2 \right)\\
        a^2A\left( \lambda_{1\tau}q_- + a\lambda_{2\tau}^*q_+^2 \right) && c_{1z}cq_z + c'_{2\perp}a^2q_+q_-
    \end{pmatrix},
    \label{eq:2by2H}
\end{align}
where the 3D wave vector $\mathbf{q} = (q_x, q_y, q_z)$ is defined as $\mathbf{q} = \mathbf{k} - \mathbf{k}_{\tau}$ using the coordinates $\mathbf{k}_{\tau}=(0,0,k_{\tau})$, where the one-photon resonance occurs on the $\Gamma$-$\mathrm{Z}$ line.
$q_\pm$ is equal to $q_x \pm i q_y$, and $A=\frac{E_0}{\Omega}$ is the vector potential amplitude.
The detailed derivations and coefficients are given in Appendix~\ref{ap:two-level_Hamiltonian}.

Weyl points appear when all components of the effective two-band Hamiltonian in Eq.~\eqref{eq:2by2H} are zero; thus, a solution is clearly given by $\mathbf{q}=0$. 
Additional solutions are obtained by writing $q_{\pm} = r e^{\pm i\varphi}$ with $r> 0$, and $\varphi \in \mathbb{R}$.
The condition that the off-diagonal term vanishes gives, 
\begin{equation}
    \lambda_{1\tau}e^{i\varphi} + ar\lambda_{2\tau}e^{-2i\varphi} = 0,
\end{equation}
whose solution for $r$ is 
\begin{equation}
    r=\frac{1}{a}\left| \frac{\lambda _{1\tau}}{\lambda _{2\tau}}\right|, \label{eq:rthreefold}
\end{equation}
since $r$ is real.
Therefore, we obtain 
\begin{equation}
    \varphi_l = \frac{1}{3}\arg{\left(-\frac{\lambda_{2\tau}}{\lambda_{1\tau}}\right)} + \frac{2l\pi}{3}
    \label{eq:threefold}
\end{equation}
for $l=0,1,2$.
The diagonal term also determines the $q_z$ coordinate of the remaining three Weyl points as
\begin{align}
    q_z=-\frac{1}{c}\frac{c_{2\perp}'}{c_{1z}}\left| \frac{\lambda _{1\tau}}{\lambda _{2\tau}}\right| ^2.
    \label{eq:zthreefold}
\end{align}
Eq.~\eqref{eq:threefold} reflects the $C_{3z}$ symmetry of the Bi$_2$Se$_3$ lattice model even under the CPL irradiation.

These results are consistent with the emergence of the Floquet double-Weyl points in the isotropic continuum Dirac model~\cite{HiraiPRR2024}, where a double-Weyl point with its charge $\pm2$ splits into two $\pm1$ Weyl points under in-plane anisotropy. 
In contrast, the present lattice model preserves the total charge of $+2$ in the $k_z>0$ half of the BZ, but realizes a distinct splitting into one $-1$ Weyl point and three $+1$ Weyl points. 
From Eqs.~\eqref{eq:rthreefold}-\eqref{eq:zthreefold}, we can see that the linear terms $\lambda _{1\tau}q_{\pm}$ induce the splitting because the four Weyl points merge at $\mathbf{k}_{\tau}$ and form a double-Weyl point in the absence of these linear terms.
As discussed in Appendix~\ref{ap:two-level_Hamiltonian}, the terms stem from the $C_{3z}$ symmetry, resulting in the four Weyl points as a split double-Weyl point.
The threefold pattern of Weyl points provides a direct manifestation of the crystal symmetry in the resonant Floquet band topology, going beyond the isotropic continuum description.

\subsection{Topological phase diagram}

\begin{figure}
    \centering
    \includegraphics[width=1.0\linewidth]{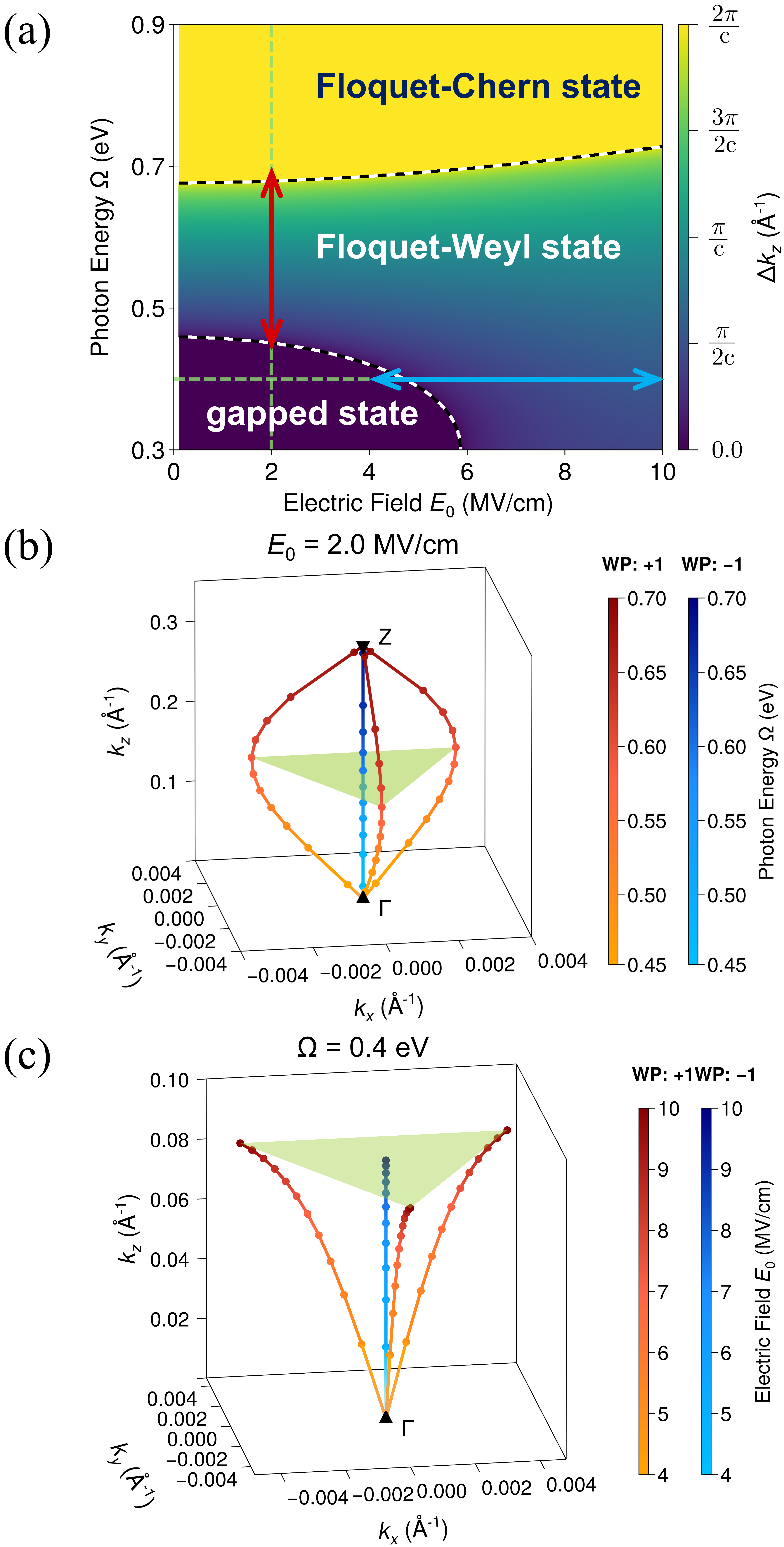}
    \caption{(a) Topological phase diagram for the photon energy $\Omega$ and the electric field $E_0$, colored by the distance of the four Weyl pairs in the $k_z$ direction, $\Delta k_z$, which contains the gapped state ($\Delta k_z = 0$), the Floquet-Weyl state ($0 < \Delta k_z < \frac{2\pi}{c}$), and the Floquet-Chern state ($\Delta k_z = \frac{2\pi}{c}$).
    The trajectories of the Weyl points (WPs) emerging at the one-photon resonance varying (b) the photon energy $\Omega$ and (c) the electric field $E_0$, in the range of the red and blue arrows in (a), respectively.
    The reddish (bluish) color bar is for the WPs with the charge $+1$ ($-1$), and the green triangles are guides for the eyes. 
    }
    \label{fig:4}
\end{figure}

Having established the Weyl-point structure for representative driving parameters, we next extend the analysis to the broader parameter space spanned by the electric-field amplitude $E_0$ and the photon energy $\Omega$.
Figure~\ref{fig:4}(a) shows the comprehensive topological phase diagram, where the color scale represents the separation $\Delta k_z$ between the paired Weyl points in the positive- and negative-$k_z$ halves of the BZ. 
In the parameter window considered here, the topological phase diagram is divided into the following three regions: the gapped state for small $\Omega$ and weak $E_0$, the Floquet-Weyl state for intermediate $\Omega$ or large $E_0$, and the Floquet-Chern state for large $\Omega$.
For weak fields and low photon energies, the quasienergy spectrum remains fully gapped, and no Weyl point is generated; thereby, the region is colored in dark with $\Delta k_z = 0$. 
Upon increasing $E_0$ or $\Omega$, pairs of Weyl points are created, leading to a Floquet-Weyl state characterized by a finite separation $\Delta k_z$. 
At still larger $\Omega$, $\Delta k_z$ rapidly increases, and the Weyl points reach the BZ boundary and pair-annihilate, yielding the Floquet-Chern state in which the 2D slices of the BZ acquire a finite Chern number $-2$ over the entire range of $k_z$.

Figures~\ref{fig:4}(b) and \ref{fig:4}(c) illustrate the trajectories of the Weyl points along two representative cuts of the phase diagram in Fig.~\ref{fig:4}(a), fixing $E_0=2.0$~MV/cm (red arrow) and $\Omega=0.4$~eV (blue arrow), respectively. 
In Fig.~\ref{fig:4}(b), increasing $\Omega$ changes the system from the gapped state to the Floquet-Weyl state, and then to the Floquet-Chern state. 
At the first phase boundary where the one-photon resonance occurs, the Weyl points are created at the $\Gamma$ point; they subsequently move toward the $\mathrm{Z}$ point as $\Omega$ increases, keeping the $C_{3z}$ symmetry as denoted by the green triangle in Fig.~\ref{fig:4}(b). 
At the second phase boundary, they annihilate at the $Z$ point, completing the topological transition to the Floquet-Chern state.
For fixed $\Omega=0.4$~eV in Fig.~\ref{fig:4}(c), increasing $E_0$ instead creates Weyl points at the $\Gamma$ point and enlarges the distance $\Delta k_z$ at least in the parameter range shown here. 
These trajectories not only provide a direct visualization of the pair creation, separation, and pair annihilation of the Weyl points in momentum space, but also can be regarded as possible paths traced under a temporal modulation of the driving parameters $E_0$ and $\Omega$.

\section{Anomalous Hall responses}
\label{sec:transport}

\subsection{Photo-induced anomalous Hall conductivity}
\label{sec:ahc}

We next discuss the transport properties of the one-photon-resonant Floquet state.
The CPL breaks time-reversal symmetry $\mathcal{T}$ and lifts the spin degeneracy of the Floquet bands, thereby generating a finite Berry curvature even in the absence of static magnetization.
To evaluate the photo-induced AHC, we use the Kubo formula for the Floquet eigenstates~\cite{OkaPRB2009},
\begin{equation}
    \label{eq:AHC}
    \sigma_{xy} = \frac{e^2}{\hbar} \sum_{n \neq m} \int_{\text{BZ}} \frac{d\mathbf{k}}{(2\pi)^3} f_{n}\Omega^{z}_{n},
\end{equation}
where 
\begin{equation}
    \label{eq:BC}
    \Omega^{z}_{n} = 2\mathrm{Im}\sum_{m\neq n}\frac{\bra{n}\partial_{k_x}\mathcal{H}_\mathrm{eff}\ket{m}\bra{m}\partial_{k_y}\mathcal{H}_\mathrm{eff}\ket{n}}{(\varepsilon_{n}-\varepsilon_{m})^2}
\end{equation}
is the time-averaged Berry-curvature of the $n$th Floquet band; $\varepsilon_{n}$ is the eigenenergy of the one-photon resonant Floquet Hamiltonian $\mathcal{H}_\mathrm{eff}$ in Eq.~\eqref{eq:H_opr}, denoted as $E - E_\mathrm{F}$ in the band structures.
For the nonequilibrium distribution function $f_n$, we employ the sudden approximation~\cite{OkaIOP2011} given as, 
\begin{equation}
\label{NEDF}
    f_{n} = \sum_{\alpha} \frac{|\Braket{\alpha|n}|^2}{1+\exp\left(\frac{\varepsilon_{\alpha} - \mu}{T}\right)},
\end{equation}
where $\ket{\alpha}$ is the eigenstate of the $\alpha$th $0$-photon band in equilibrium shown in Fig.~\ref{fig:2}(a): in the absence of the electric-field amplitude, at $E_0 = 0$, $f_n$ corresponds to the equilibrium Fermi-Dirac distribution function for temperature $T$ and the chemical potential $\mu$.
In the following calculations, we set $T = 0$ and $\mu = 0$, while we will discuss the $\mu$ dependence in the next subsection.

\begin{figure}
    \centering
    \includegraphics[width=1.0\linewidth]{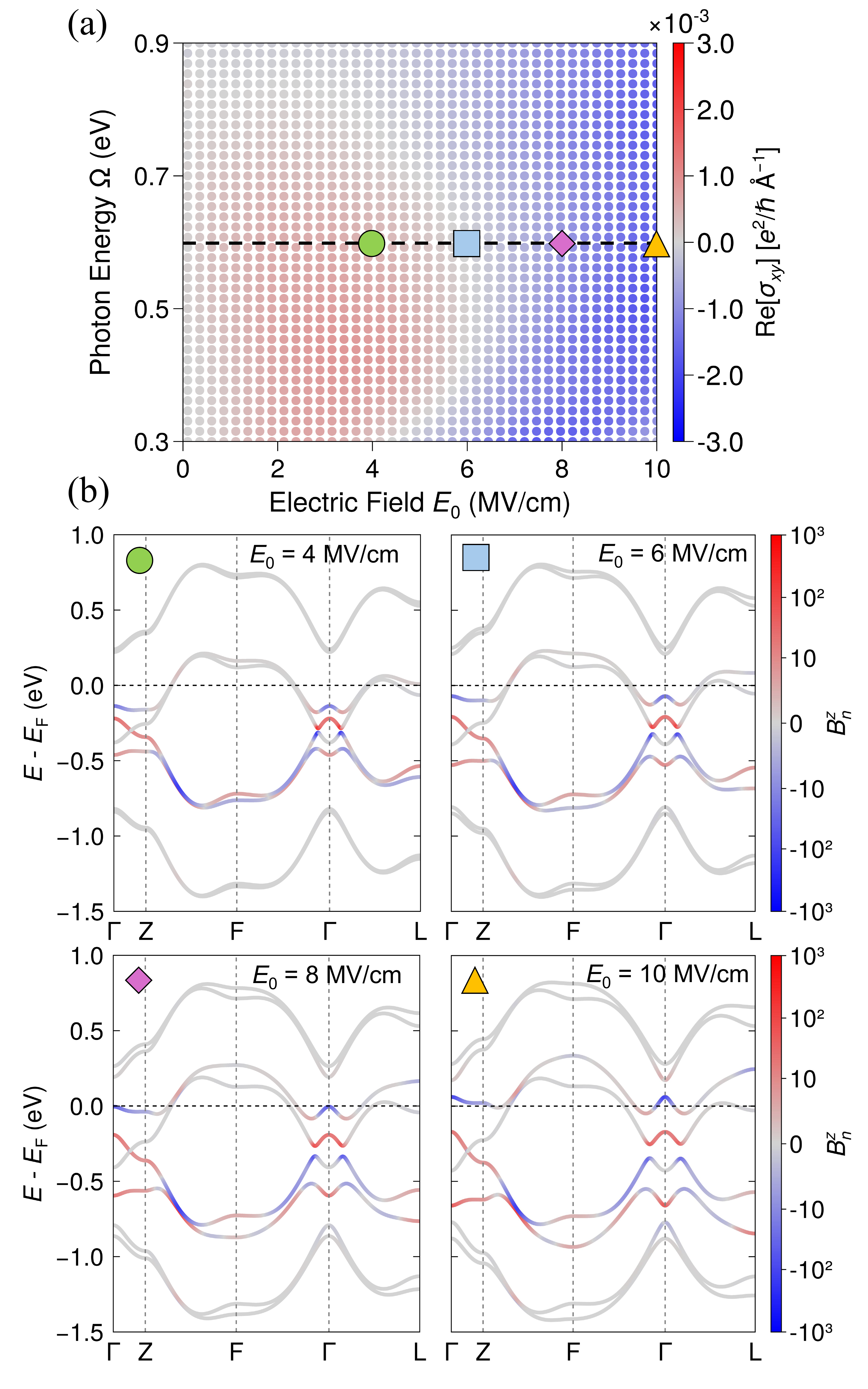}
    \caption{
    (a) Color plot of the anomalous Hall conductivity $\sigma_{xy}$ for the electric field $E_0$ and photon energy $\Omega$.
    (b) Floquet band structures with the log-scale color map of the Berry curvature distribution, $B^z_n$, in the one-photon resonant bands at $\Omega = 0.6$~eV for various electric fields denoted by the markers in (a).
    }
    \label{fig:5}
\end{figure}

Figure~\ref{fig:5}(a) shows the photo-induced AHC $\sigma_{xy}$ as a function of the electric-field amplitude $E_0$ and photon energy $\Omega$, plotted over the same parameter range as the topological phase diagram in Fig.~\ref{fig:4}(a). 
A finite anomalous Hall response appears over a broad region; however, the distribution of $\sigma_{xy}$, including the sign change around $E_0 \simeq 6$~MV/cm, does not closely follow the phase boundaries between the topologically distinct states. 
This weak correlation arises because the Weyl points are generated well below the original Fermi level in the present parameter regime, so that the Berry fluxes associated with Weyl points of opposite topological charges are almost canceled out in the BZ integral, even after weighting by the nonequilibrium distribution function.

The microscopic origin of the sign change can instead be understood from the Berry-curvature distribution $B_n^z=f_n\Omega_n^z$ shown in Fig.~\ref{fig:5}(b). 
In the weak-$E_0$ regime, $\sigma_{xy}$ is mainly governed by the positive contribution from the off-resonant band splitting around the $\Gamma$ point, which is well described by HFE discussed in Appendix~\ref{ap:HFE}. 
As $E_0$ increases, the one-photon hybridization becomes stronger and substantially modifies the band structure near the $\Gamma$ point. 
In particular, the direct gap of the parent TI bands is reduced, enhancing a negative Berry-curvature contribution around the $\Gamma$ point in the vicinity of the original Fermi level $E_\mathrm{F}$. 
This contribution eventually overcomes the positive off-resonant contribution, accounting for the sign reversal of $\sigma_{xy}$ around $E_0 \simeq 6$~MV/cm.

\subsection{Hole Doping}
\label{sec:Hole_doping}
The above results show that, when the chemical potential is kept at the original Fermi level of Bi$_2$Se$_3$, the direct contribution from the one-photon-resonant Weyl points to $\sigma_{xy}$ is largely masked by the Berry curvature distributed over the occupied Floquet bands. 
This is because the Weyl points are generated deep below the original Fermi level at the Floquet zone boundary $E - E_\mathrm{F} \simeq -\frac{\Omega}{2}$. 
To extract their topological contribution to AHE more directly, we here examine the chemical potential dependence of the AHC, especially assuming a hole-doped regime in which the Fermi level is shifted toward the resonant point.

\begin{figure}
    \centering
    \includegraphics[width=1.0\linewidth]{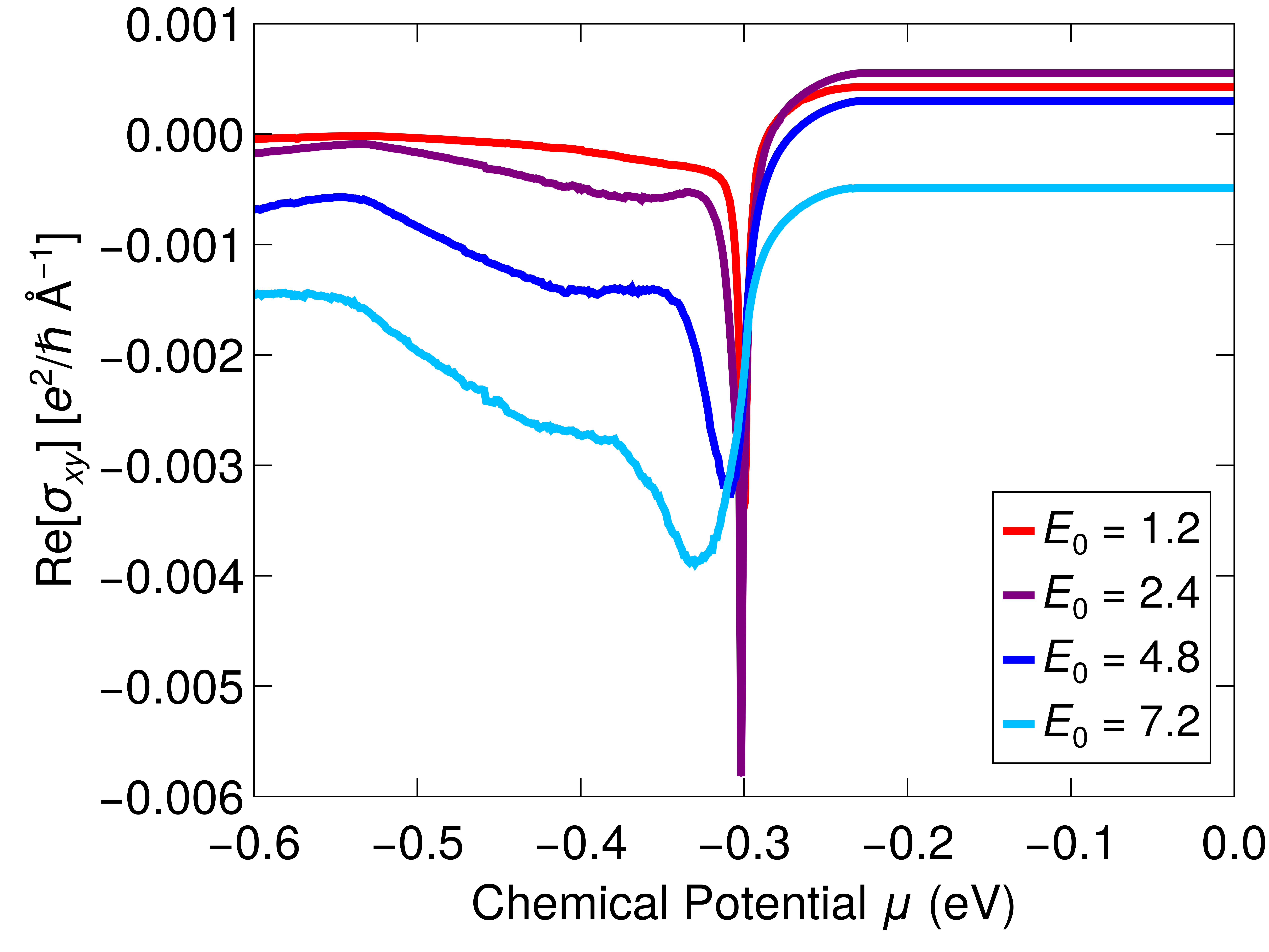}
    \caption{
    Chemical potential $\mu$ dependence of the anomalous Hall conductivity $\sigma_{xy}$ calculated by the one-photon-resonant model at $\Omega = 0.6$~eV for several electric fields $E_0$.
    }
    \label{fig:6}
\end{figure}

Figure~\ref{fig:6} shows $\sigma_{xy}$ as a function of the chemical potential $\mu$ for a fixed photon energy $\Omega = 0.6$~eV. 
A pronounced peak appears around $\mu \simeq -\frac{\Omega}{2} = -0.3$~eV, which coincides with the quasienergy at which the one-photon-resonant Weyl points emerge. 
This demonstrates that tuning the chemical potential to the Weyl-point energy enhances their contribution to the AHC. 
The peak structure is particularly sharp in the weak-$E_0$ regime since the background contribution from off-resonant band splitting remains small. 
With increasing $E_0$, the HFE contribution grows and modifies the overall Berry-curvature background, causing the peak to broaden and its maximum to shift slightly from the Weyl-point energy. 
We conclude that hole doping practically provides a useful way to distinguish the resonant Weyl-point contribution from the off-resonant HFE contribution to the AHC.

\section{Conclusion and Outlook}
\label{sec:conclusion}
In this work, we theoretically investigated the Floquet band topology and anomalous Hall response of Bi$_2$Se$3$-type three-dimensional topological insulators driven by circularly polarized light.
By focusing on the one-photon resonance between the original valence and photo-dressed conduction bands, we demonstrated the generation of the Floquet-Weyl points even for an insulating state in equilibrium. 
This result indicates that the resonant Floquet-Weyl engineering is not restricted to three-dimensional Dirac semimetals~\cite{ebiharaPRB2016,BucciantiniPRB2017,Hirai_arXiv,YoshikawaPRB2025}, but can be extended more broadly to spin--orbit coupled quantum materials.
In contrast to the isotropic continuum Dirac model~\cite{HiraiPRR2024}, where the resonant nodes host Weyl charge $\pm2$, the lattice model exhibits a characteristic threefold pattern of resonant Weyl points, reflecting the $C_{3z}$ symmetry of the crystalline structure.
By tracking the creation, motion, and annihilation of these Weyl points preserving the symmetry, we constructed a topological phase diagram consisting of the gapped, Floquet-Weyl, and Floquet-Chern states.
This symmetry-adapted resonant mechanism, therefore, offers a flexible route to designing more exotic nonequilibrium topological states.
For example, extending this idea to systems with noncrystallographic rotational symmetries, such as quasicrystals, could open an avenue for Floquet topological defects with fivefold symmetry.

We also clarified how this resonant Floquet topology manifests characteristic transport. 
A finite anomalous Hall conductivity is induced by an irradiation of circularly polarized light through time-reversal-symmetry breaking, but the response is highly nonlinear in the electric-field amplitude and even exhibits a sign reversal. 
This behavior cannot be understood from the off-resonant high-frequency expansion alone; rather, it reflects the Berry-curvature redistribution caused by one-photon hybridization, which becomes prominent when the direct gap of the parent topological-insulator bands is reduced. 
More importantly, we showed that the resonant Weyl-point contribution can be selectively enhanced by hole doping. 
By lowering the chemical potential toward the Floquet-zone boundary, i.e., half of the photon energy, the one-photon-resonant Weyl points can be placed at the Fermi level, giving rise to a pronounced peak in anomalous Hall conductivity. 
This proposes a practical way of converting the deeply located resonant Weyl points into an experimentally accessible transport signal, since hole doping of Bi$_2$Se$_3$ has already been demonstrated, for example by Ca substitution that shifts the Fermi level by about $0.3$--$0.4$~eV~\cite{Hor_hole_dope}.

More broadly, the resonant mechanism discussed here provides a highly tunable route to Floquet Weyl engineering. 
Since the energy of the one-photon-resonant Weyl points is controlled by the incident photon energy, one can, in principle, create Weyl points at selected energy scales simply by changing the laser frequency. 
This tunability is particularly useful because different nonequilibrium topological states can be explored within a single material platform, enabling systematic and quantitative comparisons of their topological responses without changing the underlying compound.
An important future direction is to clarify the surface states associated with the resonantly generated Weyl points, especially the possible emergence of Fermi-arc states connecting their surface projections~\cite{xuSci2015}. 
It is also natural to investigate exotic optical and transport responses associated with the Weyl-point Berry curvature, including nonlinear photocurrents~\cite{wuNatPhys2017,deNat2017,maNat2017}. 
Finally, while the present study focused on the hybridization between the original bands and their Floquet replicas, resonances between different Floquet-replica bands offer another promising direction. 
Such replica-replica hybridization could give more direct evidence of Floquet-band generation and help disentangle intrinsic Floquet responses from photocurrent-related contributions to the light-induced anomalous Hall effect~\cite{satoPRB2019,mciverNP2020,MurotaniPRL2023}.

\begin{acknowledgments}
This work was supported by JSPS KAKENHI (Grants No.~JP22K13998, No.~JP23K25816, No.~JP24K00586, and No.~JP26H00634) and JST PREST(No.~JPMJPR2595).
K.U. and R.O. contributed equally to this work.
\end{acknowledgments}

\appendix
\section{Details of the tight-binding model and Floquet Hamiltonian}
\label{ap:model}

In this Appendix, we summarize the details of the tight-binding Hamiltonian in Eq.~\eqref{eq:H_tb} and its Floquet components in Eq.~\eqref{eq:H_opr} used in the main text.
For the equilibrium state, we use the four-band tight-binding model for Bi$_2$Se$_3$, constructed by Mao~{\it et al.}~\cite{MaoPRB2011}.
As shown in Fig.~\ref{fig:1}(c) in the main text, the Hamiltonian is written in terms of the intra- and inter-layer nearest-neighbor hoppings associated with $\mathbf{a}_l = \left(a\cos{\phi_l}, a\sin{\phi_l}, 0\right)$ and $\mathbf{b}_l = \left(-b\sin{\phi_l}, b\cos{\phi_l}, c\right)$, respectively, where $\phi_l=\frac{2\pi}{3}(l-1)$ and $b=\frac{a}{\sqrt{3}}$.
In the numerical calculations throughout this work, we set the lattice constants to $a=4.076~\mathrm{\AA}$ and $c=9.943~\mathrm{\AA}$.

\begin{table}[b]
    \centering
    \caption{Fitted parameters of the tight-binding model and the Fermi energy with a unit of eV}.
    \label{tab:fitting_parameters}
    \begin{tabular}{cc} \hline \hline
        parameter & value (eV) \\ \hline
        $A_0$    & $4.270 \times 10^{-5}$ \\
        $B_0$    & $3.433 \times 10^{-5}$ \\
        $A_{11}$ & $0.0710$ \\
        $B_{11}$ & $0.0473$ \\
        $A_{12}$ & $0.2557$ \\
        $B_{12}$ & $0.0481$ \\
        $A_{14}$ & $0.1688$ \\
        $B_{14}$ & $0.0107$ \\
        $m_{11}$ & $-0.4799$ \\ \hline
        $E_\mathrm{F}$    & $0.0676$ \\ \hline \hline
    \end{tabular}
\end{table}

The components of the Bloch Hamiltonian in Eq.~\eqref{eq:H_tb} are decomposed $h_i(\mathbf{k}) = \sum_{l=1}^3 h_{il}(\mathbf{k})$ with
\begin{align}
 h_{0l}(\mathbf{k}) &= 2A_0\cos k_{a_l} + 2B_0\cos k_{b_l},\nonumber\\
 h_{1l}(\mathbf{k}) &= 2A_{14}\sin\phi_l\sin k_{a_l} + 2B_{14}\cos\phi_l\sin k_{b_l}, \nonumber\\
 h_{2l}(\mathbf{k}) &= -2B_{14}\sin\phi_l\sin k_{b_l} - 2A_{14}\cos\phi_l\sin k_{a_l}, \nonumber\\
 h_{3l}(\mathbf{k}) &= 2A_{12}\sin k_{a_l},\nonumber \\
 h_{4l}(\mathbf{k}) &= -2B_{12}\sin k_{b_l},~\text{and}\nonumber\\
 h_{5l}(\mathbf{k}) &= 2A_{11}\cos k_{a_l} + 2B_{11}\cos k_{b_l} + m_{11},
\label{eq:tb_comp}
\end{align}
where $k_{a_l}=\mathbf{k}\cdot\mathbf{a}_{l}$ and $k_{b_l}=\mathbf{k}\cdot\mathbf{b}_{l}$. 
As explained in the main text, in addition to the Fermi energy $E_\mathrm{F}$, we have nine model parameters $A_0$, $B_0$, $A_{11}$, $B_{11}$, $A_{12}$, $B_{12}$, $A_{14}$, $B_{14}$, and $m_{11}$; otherwise, $A_{nm}$ and $B_{nm}$ are set to zero.
The fitted parameters used in the calculations in the main text are listed in Table~\ref{tab:fitting_parameters}.
Since $A_0$ and $B_0$ are much smaller than the other hopping parameters $A_{nm}$ and $B_{nm}$, the scalar hopping term $h_0(\mathbf{k})\mathbb{I}_4$ is negligible in the low-energy window considered here.
Thus, the calculated band structure is approximately particle-hole symmetric, although the Hamiltonian does not have an exact particle-hole symmetry.

We next give the $m$th Floquet component under the CPL in the same $\Gamma$-matrix basis as the equilibrium Hamiltonian,
\begin{align}
    H_m &=  h_{0}^{(m)}(\mathbf{k}) \mathbb{I}_4 + \sum_{i=1}^5 h_i^{(m)}(\mathbf{k}) \Gamma_i,
\end{align}
where $h_i^{m}(\mathbf{k}) = \sum_{l=1}^3 h_{il}^{m}(\mathbf{k})$.
The Floquet components $h_{il}^{m}$ are given by replacing the hopping terms with wavenumber $k_{d_l}$ ($d=a,b$) in Eq.~\eqref{eq:tb_comp} as follows;
\begin{align}
    \cos{k_{d_l}} &\to J_0(dA)\cos{k_{d_l}},\nonumber\\
    \sin{k_{d_l}} &\to J_0(dA)\sin{k_{d_l}} 
\end{align}
for $m = 0$ and
\begin{align}
    \cos{k_{d_l}} &\to -J_1(dA)e^{-i\theta_{d_l}}\sin{k_{d_l}},\nonumber\\
    \sin{k_{d_l}} &\to J_1(dA)e^{-i\theta_{d_l}}\cos{k_{d_l}} 
\end{align}
for $m = 1$.
Here, $J_m(x)$ denotes the $m$th the Bessel function of the first kind, $\theta_{a_l} = \phi_l$, and $\theta_{b_l} = \phi_l + \frac{\pi}{2}$.

\section{High-frequency expansion}
\label{ap:HFE}
\begin{figure}
    \centering
    \includegraphics[width=1.0\linewidth]{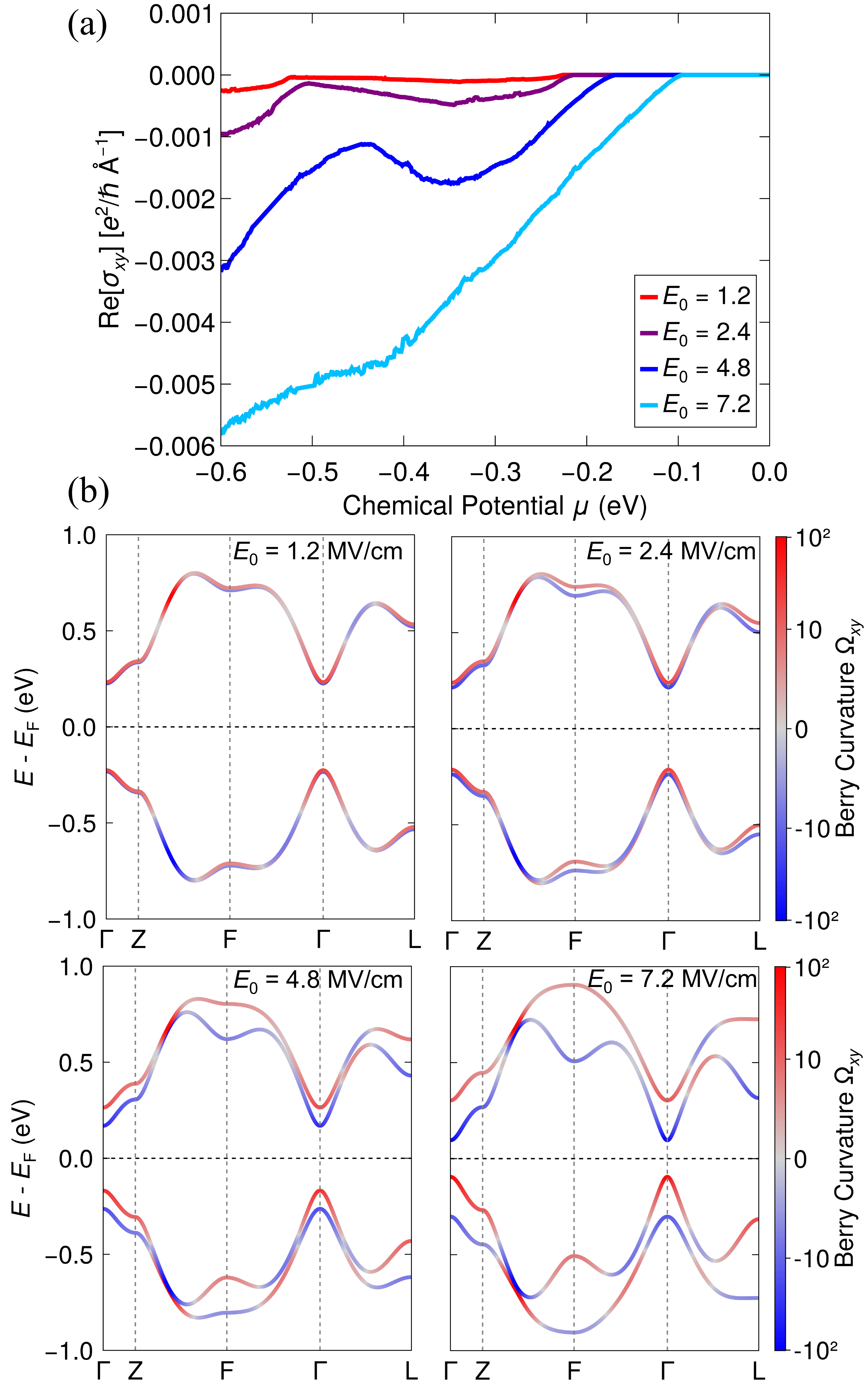}
    \caption{
    Plot analogous to Fig.~\ref{fig:6} in the main text, calculated by the effective model using high-frequency expansion.
    }
    \label{fig:7}
\end{figure}

In this Appendix, we discuss the contribution from the high-frequency limit.
We construct the effective Hamiltonian applying HFE up to the first order in $\frac{1}{\Omega}$ as
%Fig.~\ref{fig:7} shows the anomalous Hall conductivity calculated according to Eq. (\ref{eq:AHC}) using the Fermi distribution function, based on the effective Hamiltonian derived from the high-frequency expansion up to the first order:
\begin{equation}
    H_{\text{HFE}} = H_0 + \frac{[H_{-1},H_{+1}]}{\Omega},
    \label{eq:Hhfe}
\end{equation}
where $H_0$ and $H_{\pm1}$ are the Fourier components defined in Appendix~\ref{ap:model}.
We then evaluate the AHC using the same Kubo formula as in the main text, but with the eigenvalues $\varepsilon_n$, corresponding eigenstates of $H_\mathrm{HFE}$, and the Fermi-Dirac distribution function $f_n = \frac{1}{1+\exp\left(\frac{\varepsilon_n-\mu}{T}\right)}$. 
Figure~\ref{fig:7}(a) shows the chemical-potential dependence of the AHC calculated from Eq.~\eqref{eq:Hhfe}, in the same parameter setting as Fig.~\ref{fig:6}(a).

When the chemical potential lies inside the bulk gap, the AHC vanishes. 
As shown by the band structure in Fig.~\ref{fig:7}(b), the CPL-induced band splitting generates finite Berry curvature in the Floquet-engineered bands.
However, within HFE, this Berry curvature does not yield a net AHC as long as its contributions cancel after integration over the BZ. 
Upon lowering the chemical potential toward the Floquet-zone boundary at $E=-\frac{\Omega}{2}=-0.3$~eV, the sharp peaks found in the one-photon-resonant model are absent. 
This confirms that the peaks in Fig.~\ref{fig:6}(a) originate from the resonant hybridization between the original and photon-shifted bands, rather than from the off-resonant band modulation described by Eq.~\eqref{eq:Hhfe}.

Instead, the HFE model shows a broad hump structure at $\mu \sim -0.35$~eV, especially in the weak-field regime. 
This hump corresponds to the background contribution already visible in the results from the one-photon-resonant model in Fig.~\ref{fig:6}(a), and is mainly attributed to the positive Berry curvature generated near the $\Gamma$ point by the CPL-induced band splitting in Fig.~\ref{fig:7}(b). 
As the electric-field amplitude $E_0$ increases, the high-frequency contribution grows and modifies the overall chemical potential dependence: in such a high-field regime, the HFE approximation becomes less reliable.
Consequently, the high-field behavior of $H_{\rm HFE}$ no longer quantitatively agrees with that of the one-photon-resonant model.

\section{Effective two-band Hamiltonian}
\label{ap:two-level_Hamiltonian}
In this appendix, we derive the effective two-band Hamiltonian in Eq.~\eqref{eq:2by2H}.
Throughout this appendix, we set the lattice constants to unity.
First, we study one-photon-resonant points because they give rise to Floquet-Weyl points.
Next, we construct the $2\times 2$ effective Hamiltonian by expanding the lattice Hamiltonian around the resonance point.

\subsection{Resonance points}
We investigate resonance points near the $\Gamma$ point.
We expand the Hamiltonian in Eq.~\eqref{eq:H_opr} up to the first order in $A$,
and treat $H_1$ and $H_{-1}$ perturbatively below.
Before the perturbation is included, the Hamiltonian reduces to 
\begin{align}
    \mathcal{H}_{\rm eff} \simeq 
    \begin{pmatrix}
        \mathcal{H}(\mathbf{k}) & 0 \\
        0 & \mathcal{H}(\mathbf{k})-\Omega
    \end{pmatrix}. \label{eq:Hr}
\end{align}
Its eigenvalues are given by
\begin{align}
    h_0(\mathbf{k})\pm d(\mathbf{k}), \hspace{3mm}
    h_0(\mathbf{k})\pm d(\mathbf{k}) - \Omega,
\end{align}
where 
\begin{align}
d(\mathbf{k}):=\sqrt{\sum_{i=1}^5 h_i(\mathbf{k})^2}.    
\end{align}
Since one-photon resonance occurs between the valence band in the zero-photon sector and the conduction band in the $-1$-photon sector, the resonance momenta can be determined by
\begin{align}
    d(\mathbf{k}) = \frac{\Omega}{2}.
\end{align}
To obtain the momenta on the $k_z$ axis, we expand $d(\mathbf{k})$ up to first order in $k_z$ around the $\Gamma$ point:
\begin{align}
    d(\mathbf{k}) \simeq \sqrt{(6B_{12}k_z)^2+M^2},
\end{align}
where $M=m_{11}+6(A_{11}+B_{11})$.
As a result, under the condition $\Omega /2> |M|$, the resonant momenta on the $k_z$ axis are given by
\begin{align}
    k_{\tau} = \pm \frac{1}{6B_{12}} \sqrt{\frac{\Omega ^2}{4}-M^2}.
\end{align}
Therefore, the band crossing due to one-photon resonance on the $k_z$ axis occurs at
$\mathbf{k}_{\tau}=(0,0,k_{\tau})$.

We next identify the states that become resonant on the $k_z$ axis using Eq.~\eqref{eq:Hr}.
Without $H_{1}$ and $H_{-1}$, the photon sectors are block-diagonal. 
We thus calculate the eigenstates of the tight-binding Hamiltonian $\mathcal{H}(\mathbf{k}_{\tau})$,
which is given by
\begin{align}
    \mathcal{H}(\mathbf{k}_{\tau})
    = h_0(\mathbf{k} _{\tau}) \mathbb{I}_4 + \sigma _0 \otimes [h_4(\mathbf{k}_{\tau}) \tau_2 + h_5(\mathbf{k}_{\tau}) \tau _3].
\end{align}
The energy eigenvalues are $\pm d_{\tau}$, where
\begin{align}
    d_{\tau}=\sqrt{h_4(\mathbf{k}_{\tau})^2+h_5(\mathbf{k}_{\tau})^2}.
\end{align}
The eigenvectors can be analytically obtained because the spin sector is diagonal.
We denote the corresponding eigenvectors with eigenvalues $+d_{\tau}$ and $-d_{\tau}$ by $\ket{c} \otimes \ket{\sigma}$ and $\ket{v}\otimes \ket{\sigma}$, respectively. 
They are given by
\begin{align}
    \ket{c}=
    \begin{pmatrix}
        \cos \frac{\theta _{\tau}}{2} \\ i\sin \frac{\theta _{\tau}}{2} \\
    \end{pmatrix},
    \hspace{3mm}
    \ket{v}=
    \begin{pmatrix}
        i\sin \frac{\theta _{\tau}}{2} \\ \cos \frac{\theta _{\tau}}{2} \\
    \end{pmatrix},
\end{align}
where $\theta _{\tau} = \tan ^{-1}(h_4(\mathbf{k}_{\tau})/h_5(\mathbf{k}_{\tau}))$.
The one-photon resonance occurs among the valence bands in the $0$-photon sector and the conduction bands in the $-1$-photon sector.
Thus, the four degenerate states that can form the band crossing are given by
\begin{align}
\ket{c} \otimes \ket{\sigma} \otimes \ket{m=-1}, 
\hspace{3mm}
\ket{v} \otimes \ket{\sigma} \otimes \ket{m=0}, 
\label{eq:fourstates}
\end{align}
where $m$ labels the photon sector.
Hereafter, we use the shorthand notation $\ket {s, \sigma ,m}:=\ket{s}\otimes \ket{\sigma} \otimes \ket{m}$ with $s=c,v$.

Since the CPL breaks time-reversal symmetry, the fourfold degeneracy is partially lifted
after $H_{1}$ and $H_{-1}$ are included.
To see this, we focus on the subspace spanned by the four states in Eq.~\eqref{eq:fourstates}.
By performing a unitary transformation to the bases $\{ \ket{v, \uparrow, 0}, \ket{c, \downarrow, -1}, \ket{v, \downarrow, 0}, \ket{c, \uparrow, -1} \}$,
and including $H_1$ and $H_{-1}$ up to first order in $A$,
we obtain the following $4\times 4$ Hamiltonian represented by
\begin{align}
    H_{4\times 4}(\mathbf{k}_{\tau})=
    \mathcal{E}_0(\mathbf{k}_{\tau})\mathbb{I}_4
    +
    \begin{pmatrix}
        0 & 0 & 0 & 0 \\
        0 & 0 & 0 & 0 \\
        0 & 0 & 0 & Ag_{\tau } \\
        0 & 0 & Ag_{\tau }^* & 0 
    \end{pmatrix},
\end{align}
where 
$\mathcal{E}_0(\mathbf{k}) = h_0(\mathbf{k})-\Omega/2$ and
$g_{\tau}=-i[3A_{14}+\sqrt{3}B_{14}\cos k_{\tau}]$.
The states $\ket{v, \uparrow, 0}$ and $\ket{c, \downarrow, -1}$ remain degenerate with energy eigenvalue $\mathcal{E}_0(\mathbf{k}_{\tau})$; whereas the other two states are split.
Therefore, $\ket{v, \uparrow, 0}$ and $\ket{c, \downarrow, -1}$ span the 2D resonant subspace under irradiation of the CPL, giving rise to Weyl points.

\subsection{Construction of the effective Hamiltonian}
We derive the $2\times 2$ effective Hamiltonian for the two resonant states $\ket{v, \uparrow, 0}$ and $\ket{c, \downarrow, -1}$ to describe the emergent Floquet-Weyl points.
To this end, we expand the Hamiltonian around the resonance momentum $\mathbf{k}_{\tau}$ while using the eigenbasis of $\mathcal{H}_{\rm eff}(\mathbf{k}_{\tau})$.
We take the following ordered basis:
\begin{widetext}
\begin{align}
\left\{ \ket{v, \uparrow, 0}, \ket{c, \downarrow, -1}, \ket{v, \downarrow, 0}, \ket{c, \uparrow, -1},\ket{c, \downarrow, 0}, \ket{v, \uparrow, -1}, \ket{c, \uparrow, 0}, \ket{v, \downarrow, -1} \right\}.    
\end{align}
\end{widetext}
We denote the effective Hamiltonian in Eq.~\eqref{eq:H_opr} in this basis by $\mathcal{H}_{\rm eff}'$.
To derive the $2\times 2$ effective Hamiltonian, we decompose $\mathcal{H}_{\rm eff}'(\mathbf{k})$ as 
\begin{align}
    \mathcal{H}_{\rm eff}'(\mathbf{k})=
    \mathcal{E}_0(\mathbf{k})\mathbb{I}_8
    +
    \begin{pmatrix}
        H_L(\mathbf{k}) & H_{LH}(\mathbf{k}) \\
        H_{LH}^{\dagger}(\mathbf{k}) & H_H(\mathbf{k}) 
    \end{pmatrix},
\end{align}
where $H_L(\mathbf{k})$ is the $2\times 2$ block for the resonant states $\{ \ket{v, \uparrow, 0}, \ket{c, \downarrow, -1}\}$ while $H_H(\mathbf{k})$ describes the remaining six states.
The off-diagonal block $H_{LH}(\mathbf{k})$ corresponds to the hybridization between these two subspaces.

Here, we expand the Hamiltonian $\mathcal{H}_{\rm eff}'(\mathbf{k})$ around $\mathbf{k}_{\tau}$.
We use $\mathbf{q}=\mathbf{k}-\mathbf{k}_{\tau}$ and $q_{\pm}=q_x \pm iq_y$ for the expansion.
To discuss double-Weyl points and their splitting, we retain terms up to first order in $q_z$ and up to second order in $q_x$ and $q_y$.
In this approximation, $H_L$ is given by
\begin{align}
    H_L(\mathbf{q})
    \simeq
    \begin{pmatrix}
        -c_{1z}q_z - c_{2\perp} q_+q_- & A[\lambda _{1\tau} q_+ + D_3q_-^2]\\
        A[\lambda _{1\tau} q_- + D_3^*q_+^2] & c_{1z}q_z + c_{2\perp} q_+q_-
    \end{pmatrix},
    \label{eq:HL}
\end{align}
where
\begin{align}
    c_{1z}&=-6[B_{12}\cos k_{\tau}\sin \theta _{\tau}+B_{11}\sin k_{\tau}\cos \theta _{\tau}], \nonumber\\
    c_{2\perp}&=\frac{B_{12}\sin k_{\tau}}{2}\sin \theta _{\tau}
    -\frac{3A_{11}+B_{11}\cos k_{\tau}}{2}\cos \theta _{\tau}, \nonumber\\
    \lambda _{1\tau}
    &=\frac{B_{14}}{2}\sin k_{\tau}, \;\text{and}\nonumber\\
    D_3&=-\frac{3i}{8}A_{14}-\frac{i\sqrt{3}}{24}B_{14}\cos k_{\tau}.
\end{align}
For later use, we put $\delta (\mathbf{q})=c_1q_z + c_{2\perp} q_+q_-$ and $D(\mathbf{q})=A(\lambda _{1\tau} q_+ + D_3q_-^2)$, which are the diagonal and off-diagonal matrix elements of $H_L(\mathbf{q})$ in Eq.~\eqref{eq:HL}, respectively.

We note that $H_L(\mathbf{q})$ already captures band crossings that become Weyl points.
If the terms $\lambda _{1\tau}q_{\pm}$ are absent, $H_L(\mathbf{q})$ describes a double-Weyl point with topological charge $\pm 2$.
When the terms are included, the double-Weyl point is split into four Weyl points because their positions need to reflect threefold rotational symmetry $C_{3z}$ while preserving the total charge. 
This splitting generally occurs in threefold-rotationally symmetric systems because double Weyl points cannot be stabilized only by $C_{3z}$~\cite{FangPRL2012}. 

We next include the contribution from the remaining six states.
For this purpose, we employ a canonical transformation, following Ref.~\cite{YaoPRB2007}.
We introduce a matrix $M_s$, which satisfies 
\begin{align}
 M_sH_{H}-H_LM_s=H_{LH}.   
 \label{eq:Mcondition}
\end{align}
Using this matrix, we perform the transformation $\mathcal{H}_S=e^{-S}\mathcal{H}'_{\rm eff}e^{S}$ with
\begin{align}
    S=
    \begin{pmatrix}
        0 & M_s \\
        -M_s^{\dagger} & 0
    \end{pmatrix}.
\end{align}
The effective two-band Hamiltonian is obtained from the block of $\mathcal{H}_S$ projected on the 2D resonant subspace.
Near the momenta $\mathbf{k}_{\tau}$, the matrix elements of $H_L(\mathbf{q})$ are negligible compared with the energy separation in $H_H(\bm{q})$.
Thus, Eq.~\eqref{eq:Mcondition} approximately gives $M_s \simeq H_{LH}H_H^{-1}$.
By keeping up to second order in $H_{LH}$, we obtain
\begin{align}
    H_{2\times 2}(\mathbf{q})
    \simeq
    \mathcal{E}_0(\mathbf{q})\mathbb{I}_2+ H_L(\mathbf{q}) 
    -H_{LH}(\mathbf{q})
    H_H^{-1}(\mathbf{q})
    H_{LH}^{\dagger}(\mathbf{q}).
    \label{eq:SW}
\end{align}

To evaluate the term $H_{LH}(\mathbf{q})H_H^{-1}(\mathbf{q})H_{HL}(\mathbf{q})$ in Eq.~\eqref{eq:SW}, we decompose the subspace spanned by the six states.
One part consists of the two states split off by the CPL at $\mathbf{k}_{\tau}$, while the other consists of the remaining four nonresonant states.
Accordingly, we write
\begin{align}
\begin{pmatrix}
        H_L & H_{LH} \\
        H_{LH}^{\dagger} & H_H 
\end{pmatrix}
=    
\begin{pmatrix}
        H_L & H_{LD} & H_{LN} \\
        H_{LD}^{\dagger} & H_D & H_{DN} \\
        H_{LN}^{\dagger} & H_{DN}^{\dagger} & H_N
\end{pmatrix},
\end{align}
where $H_D$ is the $2 \times 2$ block for the split-off states and $H_N$ are the $4 \times 4$ block for the remaining nonresonant states.
Here, $H_{LH}=(H_{LD}~ H_{LN})$ is the $2\times 6$ block.
We now approximate $H_H^{-1}$ by neglecting $H_{DN}$, which corresponds to the hybridization between the split-off states and four nonresonant states; namely, we use
\begin{align}
    H_H^{-1}\simeq 
    \begin{pmatrix}
        H_D & 0 \\
        0 & H_N
    \end{pmatrix}^{-1}.
\end{align}
In this case, we have
\begin{widetext}
    \begin{align}
    H_{LH}(\mathbf{q})
    H_H^{-1}(\mathbf{q})
    H_{LH}^{\dagger}(\mathbf{q})
    \simeq 
    H_{LD}(\mathbf{q})
   H_D^{-1}(\mathbf{q})
    H_{LD}^{\dagger}(\mathbf{q})
    +
    H_{LN}(\mathbf{q})
    H_N^{-1}(\mathbf{q})
    H_{LN}^{\dagger}(\mathbf{q}).
    \label{eq:HHapp}
    \end{align}
\end{widetext}

Within this approximation, we keep terms up to first order in $q_z$ and up to second order in $q_x$ and $q_y$.
It is sufficient to retain only the leading terms of $q_x, q_y$ and $q_z$ to investigate their contributions.
We first evaluate the contribution from the two states split off by the CPL.
The relevant blocks are approximated as
\begin{align}
    H_D(\mathbf{q})
    &\simeq
    \begin{pmatrix}
        -\delta(\mathbf{q}) & Ag_{\tau}\\
        Ag_{\tau}^* & \delta(\mathbf{q})
    \end{pmatrix}, \\
    %H_{LD}(\mathbf{q})
    %\simeq
    %\begin{pmatrix}
    %   0 & S(\mathbf{q}) \\
    %    S(\mathbf{q})^* & 0
    %\end{pmatrix},
    H_{LD}(\mathbf{q})
    &\simeq
    \begin{pmatrix}
        0 & -iAR_{\tau}q_- \\
        iAR_{\tau}q_+ & 0
    \end{pmatrix},
\end{align}
%where $G(\mathbf{q})=Ag_{\tau}$ and $S(\mathbf{q})=-iAR_{\tau}q_-$.
where the coefficient $R_{\tau}$ is given by
\begin{align}
 R_{\tau}=-\frac{B_{12}\sin k_{\tau}}{2}\cos \theta _{\tau}
 -\frac{3A_{11}+B_{11}\cos k_{\tau}}{2}\sin \theta _{\tau}.
\end{align}
Using these expressions, we obtain
\begin{align}
    H_{LD}(\mathbf{q})H_D^{-1}(\mathbf{q})H_{LD}^{\dagger}(\mathbf{q})
    \simeq
    \begin{pmatrix}
        0 & -A\frac{R_{\tau}^2}{g_{\tau}}q_-^2 \\
        -A\frac{R_{\tau}^2}{g_{\tau}^*}q_+^2 & 0
    \end{pmatrix}. 
    \label{eq:HDcont}
\end{align}

We also study the contribution from the four nonresonant states.
The block $H_N$ is approximated as
\if0
\begin{align}
        H_N(\bm{q})
    &\simeq
    \begin{pmatrix}
        \Omega + \delta(\mathbf{q}) & G(\mathbf{q}) & 0 & -S(\mathbf{q}) \\
        G(\mathbf{q})^* & -\Omega - \delta(\mathbf{q}) & -S(\mathbf{q})^* & 0 \\
        0 & -S(\mathbf{q}) & \Omega + \delta (\mathbf{q}) & D(\mathbf{q}) \\
        -S(\mathbf{q})^* & 0 & D^*(\mathbf{q}) & -\Omega -\delta (\mathbf{q})
    \end{pmatrix}, \label{eq:HN}
\end{align}
\fi
\begin{align}
        H_N(\bm{q})
    &\simeq
    \begin{pmatrix}
        \Omega + \delta(\mathbf{q}) & Ag_{\tau} & 0 & iAR_{\tau}q_- \\
        Ag_{\tau}^* & -\Omega - \delta(\mathbf{q}) & -iAR_{\tau}q_+ & 0 \\
        0 & iAR_{\tau}q_-  & \Omega + \delta (\mathbf{q}) & D(\mathbf{q}) \\
        -iAR_{\tau}q_+ & 0 & D^*(\mathbf{q}) & -\Omega -\delta (\mathbf{q})
    \end{pmatrix}. \label{eq:HN}
\end{align}
We further decompose $H_N$ and $H_{LN}$ as
\begin{align}
    H_N = 
    \begin{pmatrix}
        H_a & H_{ab} \\
        H_{ab}^{\dagger} & H_b
    \end{pmatrix},
    \hspace{3mm}
    H_{LN}=
    (H_{La}, H_{Lb}).
\end{align}
Here, $H_a, H_b$ and $H_{ab}$ are $2\times 2$ blocks defined by Eq.~\eqref{eq:HN}, 
The corresponding blocks of $H_{LN}$ are
\begin{align}
    H_{La}(\mathbf{q})
    &\simeq
    \begin{pmatrix}
        g_{\tau}^*q_- & A(\eta-c_{2\perp})q_- \\
        A(\eta +c_{2\perp})q_+ & g_{\tau}q_+
    \end{pmatrix}, \\
    H_{Lb}(\mathbf{q})
    &\simeq
    \begin{pmatrix}
        -i\kappa q_z-iR_{\tau}q_+q_- & 0 \\
        0 & i\kappa q_z+iR_{\tau}q_+q_-
    \end{pmatrix}, 
\end{align}
where $\eta = -(3A_0+B_0\cos k_{\tau})/2$ and $\kappa =6B_{12}\cos k_{\tau}\cos \theta _{\tau}-6B_{11}\sin k_{\tau}\sin \theta _{\tau}$.
Here, we approximate $H_{N}^{-1}$ by separating $H_N$ into its block-diagonal part and off-diagonal part.
We define
\begin{align}
    M_N=
    \begin{pmatrix}
        H_a & 0 \\
        0 & H_b
    \end{pmatrix},
    \hspace{3mm}
    V_N=H_{N}-M_N.
\end{align}
Because we have
\begin{align}
    H_N^{-1}=M_N^{-1}-M_N^{-1}V_NM_N^{-1}+\cdots,
\end{align}
we substitute this expansion for $ H_{LN}(\mathbf{q})H_N^{-1}(\mathbf{q})H_{LN}^{\dagger}(\mathbf{q})$.
The terms with $V_N$ give higher-order corrections in the present expansion.
Therefore, we approximate $H_N^{-1}\simeq M_N^{-1}$.
Using this approximation, we obtain
\begin{align}
    H_{LN}H_N^{-1}H_{LX}^{\dagger}
    \simeq
    H_{La}H_a^{-1}H_{La}^{\dagger}
    +
    H_{Lb}H_b^{-1}H_{Lb}^{\dagger}.
\end{align}
The second term is beyond the order retained in the present approximation, and therefore we neglect it.
Thus, the leading contribution from the nonresonant states is given by the first term.
A straightforward calculation gives
\begin{widetext}
\begin{align}
    H_{La}(\mathbf{q})H_a^{-1}(\mathbf{q})H_{La}^{\dagger}(\mathbf{q})
    \simeq
    \begin{pmatrix}
        \frac{|g_{\tau}|^2}{\Omega}q_+q_- & -Ag_{\tau}\left[ \frac{2c_{2\perp}}{\Omega}+\frac{|g_{\tau}|^2}{\Omega ^2} \right] q_-^2\\
        -Ag_{\tau}^*\left[ \frac{2c_{2\perp}}{\Omega}+\frac{|g_{\tau}|^2}{\Omega ^2} \right] q_+^2
        & -\frac{|g_{\tau}|^2}{\Omega}q_+q_-
    \end{pmatrix}. 
    \label{eq:HNcont}
\end{align}
\end{widetext}
By defining
\begin{align}
    &c_{2\perp}'=c_{2\perp} + \frac{|g_{\tau}|^2}{\Omega}, \\
    &\lambda _{2\tau}=\frac{R_{\tau}^2}{g_{\tau}}+D_3+g_{\tau}\left[ \frac{2c_{2\perp}}{\Omega}+\frac{|g_{\tau}|^2}{\Omega ^2} \right],
\end{align}
and combining Eqs.~\eqref{eq:HDcont} and \eqref{eq:HNcont} with Eq.~\eqref{eq:HL}, we obtain the two-band Hamiltonian in Eq.~\eqref{eq:2by2H}.

Finally, we discuss a relationship between the effective two-band Hamiltonian and $C_{3z}$ symmetry.
Since we have expanded the lattice Hamiltonian around the resonance point on the $k_z$ axis, the effective Hamiltonian should respect this $C_{3z}$ symmetry.
Although Floquet double-Weyl points emerge in the isotropic continuum Dirac model~\cite{HiraiPRR2024}, in our effective two-band Hamiltonian, the off-diagonal element contains terms linear in the in-plane momentum $q_{\pm}$, namely $\lambda _{1\tau}q_{\pm}$, which are absent in the isotropic Dirac model.
The presence of the linear terms $\lambda _{1\tau}q_{\pm}$, which split the double-Weyl point, can be understood from the $C_{3z}$ symmetry.
Under the threefold rotation, $q_-^2$ transforms as $q_-^2 \rightarrow e^{i2\pi/3}q_-^2$
and $q_{+}$ also transforms as $q_+ \rightarrow e^{i2\pi/3}q_+$.
This means that $q_-^2$ and $q_+$ transform in the same way under $C_{3z}$ symmetry.
Therefore, the two terms are allowed to coexist in the same off-diagonal element under the $C_{3z}$ symmetry.
In contrast, because the continuum Dirac model has continuous rotational symmetry,
$q_-^2$ and $q_+$ transform differently under rotation, and therefore do not coexist in the same off-diagonal component.
Thus, the emergent linear terms are allowed by the $C_{3z}$ symmetry of the lattice model, reflecting the crystal structure of Bi$_2$Se$_3$.

%\nocite{*}

\bibliography{FW_260701}
\end{document}